\documentstyle[12pt]{article}

\def\lromn#1{\uppercase\expandafter{\romannumeral#1}}

\begin{document}

\begin{flushright}
TU/97/523 \\
\end{flushright}

\vspace{0.5cm}

\begin{center}
\begin{large}

\renewcommand{\thefootnote}{\fnsymbol{footnote}}
\bf{
Elementarity in Open Systems \\
- {\em Unstable Particle Decay in Medium} -
}
\footnote[2]
{
Talk given at AIJIC97 on Recent Developments in Nonperturbative Quantum
Field Theory , held at Seoul, May 26-30, 1997.
To appear in the Proceedings (World Scientific).
}
\end{large}

\vspace{1cm}
\begin{large}
M. Yoshimura

Department of Physics, Tohoku University\\
Sendai 980-77 Japan\\
\end{large}

\end{center}

\begin{center}
\vspace{1.5cm}

{\bf ABSTRACT}
\end{center}

The problem of unstable particle decay is discussed to show how
elementarity of a subsystem immersed in an infinitely larger environment
is lost. The decay law, when the same kind of particles as decay product 
make up a thermal medium, is worked out in detail.
The relic abundance of unstable particles does not suffer from the Boltzmann
suppression crucial at very low temperatures, 
because the off shell contribution not considered in the Boltzmann approach, 
becomes dominant at low temperatures.
The short-time behavior of the non-decay probability is also clarified,
which is important to discuss physical relevance of the non-observation
of nucleon decay.
Two powerful methods in this respect are the operator and the path
integral approach, both of which are reviewed.


\newpage

\begin{center}
\begin{large}
{\bf \lromn 1 Introduction}
\end{large}
\end{center}

Ideal elementary processes rarely occur. They presumably never occur
except in carefully prepared laboratory experiments.
It might be useful to recall that even processes usually considered 
elementary should actually be regarded  taking place in complex environments.
This has to do with how one separates a system in question from a
surrounding environment.
For instance, the beta decay, the fundamental weak process 
\( \:
n \:\rightarrow  \: p + e + \bar{\nu} _{e}
\: \),
when it occurs in nuclei, is compounded by effects of nuclear strong 
interaction with the rest of nucleons.
The most spectacular of this kind is how nucleon decay proceeds.
The extremely weak process of baryon number violating process at the
quark-lepton level must inevitably occur in the hadronic environment,
a nucleon.
A repeated question on the nucleon decay raised in the past is 
how the extremely slow process
of the lifetime of order $10^{31}$ years or even larger is modified
by strong interaction having the time scale of order, $10^{-23}$ seconds.
In this respect one must deal with the short time limit of the decay law, and
one must face the fact that the exponential decay law is not
exact in quantum mechanics.\cite{non-exponential decay law}

In many cases of analysis on a physical system,
one focusses on a small subsystem put in an infinitely large closed system.
Behavior of the subsystem then exhibits an apparent loss of elementarity due to
interaction with the larger environment over which we have no control.
Phrased somewhat differently, one often asks
how a quantum system evolves with time under influence of an environment 
described by some mixed state.
When one has a clear idea of how to separate the small system 
in question from the environment,
the essential part of our problem becomes how elementary processes are
modified in dissipative medium.
Dissipation reflects the fact that we have ignorance of the larger
environment.
We believe that the problem should be analyzed from the first
principles of quantum mechanics, at least at the conceptual level.
In condensed matter physics the problem of this kind is generally known as
quantum Brownian motion or quantum dissipation.

It has become increasingly clear to us that the bulk of the past works
on quantum dissipation relies on the simplified form of quantum friction,
namely dissipation which is local in time.
This corresponds to the exponential decay law when one examines
the fate of some initial excitation including the unstable particle.
This approximation is excellent in the most dominant phase of
dissipation.
But it fails both in early time and very late time behavior of
the quantum system immersed in dissipative medium.
Some formalism exists to deal with more general nonlocal dissipation, but
they are not very useful to the problems we would like to address.
Indeed, some of our new results have not been recognized in the past.

The basic idea taken by most approaches in 
the past \cite{q-dissipation in hos}, and also the one
I use here, is that existence of
a continuously infinite number of environment variables 
coupled to a finite number of 
subsystem variables is the essential part of dissipation.
In this view the origin of dissipation is that once something flows from the
small system to the infinite environment, then it practically never returns
to the small system.
Dissipation, or at least something recognized as
such, occurs when one does not and cannot make a 
measurement on the environment.
It is well known that even for a pure quantum system the entropy, a measure
of dissipation, is nonvanishing once one traces out a part of the quantum
system.\cite{srednicki 93}
Thus one postulates that detailed modeling of the environment and
its coupling to the small system should be unimportant to the
dissipative behavior of a small system.
After integrating out environment variables, one should have
only a few phenomenological parameters to describe the dissipation,
ultimately obtainable from experiments.
Despite this phenomenological nature one should base all auguments
on rigorous quantum physics, and modeling is inevitable.

The simplest, yet the most fundamental model of quantum dissipation
is harmonic oscillator coupled to infinitely many oscillators that
make up a bath in a mixed state. 
As an ideal limit of the mixed state one can also consider a pure
quantum state.
The nonlinearity not considered here
is presumably important to restore the complete
equilibrium of the small plus the environment system.
But our interest is solely in the behavior of the small system, which is
modified by interaction with a larger environment, and
we are not interested in the environment part.
For this discussion the linear approximation should be adequate.

The problem of how a small system behaves under influence 
of a larger environment
has been investigated in a variety of approaches.
Two powerful methods to analyze
this problem are the quantum Langevin equation,
and the path integral approach.\cite{qbm path review}
Both approaches have merits and demerits, but when combined, they
become very powerful.
I shall first give a fundamental result from the operator approach, which
makes clear the meaning of some basic functions frequently used later.
In the second part of this lecture I also explain the path integral approach.
Our main results are in our two papers already in print.
\cite{jmy-96-1}, \cite{jmy-96-2} 

In this lecture I shall exclusively discuss one single problem
of this rather broad subject.
It is the decay process in medium.
The problem we set up is described as follows.
Suppose that one would like to know how the decay of unstable particles
proceeds when they decay into two particles that also make up the bulk
of the environment.
It is not clear how the decay law known in the pure quantum case
persists or is modified. 
For instance, is the well known time 
dependence \cite{non-exponential decay law},
namely the exponential law followed by the power law decay modified? 
An indication of substantial modification to this decay law
is suggested by the following consideration.
Some amount of parent particles is clearly left behind in medium, 
even much later
than the decay lifetime, because in thermal environment even a heavier
parent particle may be created by energetic daughter particles of
smaller mass.
If this is the case, 
what is the fraction of the parent particles left behind?
We would like to answer these questions and elucidate the basic law 
by using a rigorous result of the general formalism.
I shall first give a fairly self-contained discussion of the general 
formalism, because in some literature an unnecessary approximation of
the local friction is made at the level of the general formalism, 
obscuring the validity of results obtained when 
the general formalism is applied to some specific problem.
Our result applied to the unstable particle decay in thermal medium
indeed casts a serious doubt on the familiar result
based on the Boltzmann equation that uses quantities on the mass shell.

\vspace{0.5cm} 
\begin{center}
\begin{large}
{\bf \lromn 2 The Model}
\end{large}
\end{center}

In the present approach 
one models the environment by a continuously infinite set of
harmonic oscillators of some arbitrary spectrum and couples it to
the subsystem via a bilinear term.
As will be explained below, the two particle states of decay 
product may be taken to be the environment oscillator of this kind.
Let the subsystem variable in question be denoted by $q$ and 
the environment variable by $Q({\omega })$. 
For simplicity we assume that the subsystem has one degree of freedom,
but it should be evident to extend it to any finite number of
degrees of freedom.
The Lagrangian of our problem consists of three parts: 
\begin{equation}
L = L_{q}[q] + L_{Q}[Q] + L_{{\rm int}}[q\,, Q] \,.
\end{equation}
We take for the system-environment interaction the bilinear term:
\begin{eqnarray}
&&
L_{q} = \frac{1}{2}\, \dot{q}^{2} - V(q) \,, 
\\ &&
L_{Q} = 
\frac{1}{2}\, \int_{\omega _{c}}^{\infty }\,d\omega \,\left( \,
\dot{Q}^{2}(\omega ) - \omega ^{2}\,Q^{2}(\omega ) \,\right)
\,,
\\ &&
L_{{\rm int}} = 
-\,q\,\int_{\omega _{c}}^{\infty }\,d\omega \,c(\omega )Q(\omega )
\,.
\end{eqnarray}
Here $\omega _{c}$ taken to be positive is the smallest of the environment
frequency spectrum, and $c(\omega )$ describes the strength distribution
of environment-system interaction.
In the present work we only consider the harmonic oscillator for the
small system:
\begin{equation}
V(q) = \frac{1}{2}\, \omega _{0}^{2}\,q^{2} \,.
\end{equation}

The model of environment including its interaction to the subsystem 
is characterized by the quantity,
\begin{equation}
r(\omega ) = \frac{c^{2}(\omega )}{2\omega } \,, 
\end{equation}
which we called the response weight.\cite{hjmy 96} 
This quantity is fundamental to the rest of our discussion.
The simplest response weight familiar in condensed matter physics
is given in terms of three parameters,
a threshold $\omega _{c}$, an index $\alpha $ and a strength $c$,
\begin{equation}
r(\omega ) = c\,(\omega - \omega _{c})^{\alpha } \,.
\end{equation}
The other hidden parameter here is the cutoff frequency, $\Omega $,
above which the response weight vanishes, or it is nonvanishing only for
\( \:
\omega _{c} < \omega < \Omega \,.
\: \)
In condensed matter physics the gapless case of $\omega _{c} = 0$ 
is especially popular, and the casses of
\( \:
\alpha = 1, \; < 1, \; > 1 \,, 
\: \)
are called the Ohmic, sub-Ohmic, and super-Ohmic dissipation.

Our main interest is in the unstable particle decay.
In this case the form of the response weight is more complicated than
those given above.
Let us first discuss how one identifys the environment variable $Q(\omega )$
when the same species of particles as the decay product 
make up the environment.
We take a relativistic field theory of Yukawa type of decay interaction:
\begin{equation}
{\cal L}_{{\rm int}} = \frac{\mu }{2}\,\varphi \,\chi ^{2} \,,
\end{equation}
where $\varphi $ is the decaying parent and $\chi $ the daughter particle,
with $\mu $ the coupling of mass dimension.
We shall use the capital letter $M$ for the mass of the parent $\varphi $ 
and $m$ for the mass of the daughter $\chi $.
The threshold of the decay in vacuum is 
\( \:
\omega_{c} = \sqrt{\vec{k}^{2} + 4m^{2}} \,,
\: \)
but this is modified in thermal environment, as will be made clear
shortly. The decaying particle, 
or the Fourier component of its field operator
\( \:
q_{\vec{k}} = \int\,d^{3}x\,\varphi(\vec{x})\,e^{i\vec{k}\cdot \vec{x}} \,, 
\: \)
couples to the two-body operator of $\chi$'s, and we identify the
environment variable as
\begin{equation}
c_{\vec{k}}(\omega )\,Q_{\vec{k}}(\omega ) = \frac{\mu }{2}\,
\int\,d^{4}x\,\chi ^{2}(x)\,e^{-\,i\vec{k}\cdot \vec{x} + i\omega x_{0}}
\,.
\end{equation}

The translational invariance makes each $\vec{k}-$mode independent.
It is crucial to realize that each momentum state of the parent particle
couples to a continuously infinite number of
two particle states of daughter particles.
The approximation, implicit here,
that two particle $\chi $ states are made of independent, non-interacting
two particle states
is equivalent to that we neglect the final state interaction of $\chi$'s.
We shall omit the vector notation such that
\( \:
\vec{k} \rightarrow k \,, 
\: \)
and when confusion does not arise, we also omit the mode $k$ altogether.

For definiteness, we take a thermal environment of temperature 
$T = 1/\beta $.
The response weight $r(\omega )$ is then calculable using the
technique of the finite temperature field theory.\cite{hjmy 96} 
It is the discontinuity or the imaginary part of the self-energy 
$\Pi (\omega \,, \vec{k})$ of $\varphi $ field in thermal
medium:
\begin{equation}
r(\omega ) = \frac{1}{2i\,\pi^{2} }\,\left( \,\Pi(\omega - i0^{+}) - 
\Pi (\omega + i0^{+}) \,\right) 
\equiv -\,\frac{1}{\pi }\,\Im \,\Pi(\omega ) \,.
\end{equation}

Let us explain some details of this calculation of the response
weight in the subthreshold region of 
\( \:
|\omega | < k \,.
\: \)
The discontinuity is readily calculable from the imaginary-time formalism
and it is given by \cite{weldon}
\begin{eqnarray}
-\,\Im \Pi\, (\omega ) &=& \frac{\mu ^{2}}{16\pi k}\,
\int_{- \omega _{-}}^{\infty }\,dE\,\left( \,n(E) - n(E + \omega )\,\right)
\,, \label{subthreshold imaginary part} \\
\omega _{\pm } &=&
\frac{\omega }{2} \pm \frac{k}{2}\,
\sqrt{\,1 - \frac{4m^{2}}{\omega ^{2}-k^{2}}\,} \,,
\end{eqnarray}
where
\begin{equation}
n(E) = \frac{1}{e^{\beta E} - 1}
\end{equation}
is the Planck distribution function of $T = 1/\beta $.
Since
\begin{equation}
n(E) - n(E + \omega ) = n(E)(\,1 + n(E + \omega )\,) -
n(E + \omega )(\,1 + n(E)\,) \,, 
\end{equation}
the imaginary part (\ref{subthreshold imaginary part}) for
$|\omega | < k$ is a sum of the two contributions,
\( \:
\chi + \varphi \rightarrow \chi 
\: \)
and its inverse process that is allowed to occur in thermal medium.
Note that $\varphi $ can be off the mass shell: 
\( \:
\omega ^{2} - \vec{k}^{2} \neq 
\: \)
the $\varphi $ mass$^{2}$.
The factor $1 + n$ represents the effect of stimulated boson emission.

On the other hand, for $\omega > \sqrt{\,k^{2} + 4m^{2}\,}$ 
relevant physical processes are
\( \:
\varphi \:\leftrightarrow  \: \chi + \chi \,.
\: \)
Since
\begin{equation}
(\,1 + n(E)\,)(\,1 + n(E + \omega )\,)- n(E)n(E + \omega ) = 
1 + n(E) + n(E + \omega ) \,,
\end{equation}
the imaginary part of the self-energy is given by
\begin{eqnarray}
-\,\Im \Pi\, (\omega ) &=& \frac{\mu ^{2}}{32\pi k}\,
\int_{\omega _{-}}^{\omega _{+}}\,dE\,\left( \,1 + 2n(E)\,\right)\,,  \\
&\rightarrow &
\frac{\mu ^{2}}{32\pi }\,\sqrt{\,1 - \frac{4m^{2}}{\omega ^{2} - k^{2}}\,}
\,(\,1 + 2e^{-\,\beta \omega /2}\,) \,.
\end{eqnarray}
The last limit is valid as $\omega  \rightarrow \infty $. 
When divided by $\omega $, the first term here
\( \:
\Gamma (\omega ) = \Im \,\Pi (\omega )/\omega 
\: \)
is 
\( \:
\approx \mu ^{2}/(32\pi \,\omega ) \,, 
\: \)
which is the decay rate of $\varphi \rightarrow \chi \chi $ in
vacuum, including the effect of prolonged lifetime at
\( \:
\omega  \gg m_{\varphi } \,.
\: \)\cite{weldon} 

The result of computation is now summarized.
For $\omega > \sqrt{k^{2} + 4m^{2}}$ the response weight is \cite{jmy-96-1} 
\begin{eqnarray}
r(\omega ) = \frac{\mu ^{2}}{32\pi^{2} }\,
\left( \,\sqrt{\,1 - \frac{4m^{2}}{\omega ^{2} - k^{2}}\,} +
\frac{2}{k\beta }\,\ln \frac{1 - e^{-\beta \omega _{+}}}
{1 - e^{-\beta |\omega _{-}|}}\,\right) \,. 
\label{response for decay}
\end{eqnarray}
For $0 < \omega < k$ only the second term in the bracket of 
Eq.(\ref{response for decay}) contributes.
Note that $r(\omega ) \rightarrow $ a constant
($\approx  M/\pi \, \times $ decay rate in the rest frame of $\varphi $)
as $\omega \rightarrow \infty $.

We note that there is a gap between $k$ and $\sqrt{k^{2} + 4m^{2}}$
for which $r(\omega ) = 0$.
Existence of the gap is important in discussing the analytic property
of some basic functions that appear later.
However, the location of the gap in thermal medium differs from that
in vacuum.
The gap ceases to exist for the massless case $m= 0$.

\vspace{0.5cm} 
\begin{center}
\begin{large}
{\bf \lromn 3 Diagonalization of the total system
}
\end{large}
\end{center}

A typical question one frequently asks with regard to the system behavior
is how a pure state of the $q$ system
evolves in time in environment described by a mixed state such as
the thermal one of $Q(\omega )$'s.
We shall first answer this by giving explicit time dependent operator
solution for $q(t) \,, p(t) = \dot{q}(t)$ 
that is written in terms of initial values,
\( \:
q_{i} \,, p_{i} \,, Q_{i}(\omega ) \,, P_{i}(\omega ) \,.
\: \)
With this, one can clearly express correlators such as
\( \:
\langle q(t_{1})q(t_{2}) \rangle
\: \)
in terms of the probability distribution of these values in any initial state.
Closest to our present approach is the classic work of 
Ullersma \cite{ullersma}, with an important difference of the presence of a gap
in the environment spectrum, which yields different behaviors 
of the correlators.
The gapless case is applied  for instance to phonons in medium,
but in the case of our interest such as the unstable
particle decay in medium, a gap exists if the mass of daughter particle
is finite.

The model itself, a harmonic system coupled to an infinite number 
of harmonic oscillators,
frequently appears in many idealized physical problems.
We utilize in the present investigation the
exact solution of the Hamiltonian eigenvalue problem to this system,
which might also be useful in other contexts.
With this exact eigen operator the Heisenberg time evolution becomes evident.
A great advantage of this way of solving the present problem is that
one can employ a full analogy to the scattering problem, especially
the analyticity based on elastic unitarity.
What happens is that an isolated spectrum of the subsystem above
the two particle threshold becomes unstable due to the interaction
with the environment,
and the single particle pole enters into the second Riemann sheet
below the cut real axis.
Due to the structure of our model, the unitarity relation is
saturated by the elastic two-body state.
The system then becomes integrable.

We first consider a related problem of diagonalization of the infinite
dimensional matrix of the potential part ${\cal V}$:
\begin{equation}
V = \frac{1}{2}\, \left( \begin{array}{cc}
q & Q(\omega )
\end{array}
\right) \,{\cal V}\,
\left( \begin{array}{c}
q  \\ Q(\omega' )
\end{array}
\right) \,, \hspace{0.5cm} 
{\cal V} = \left( \begin{array}{cc}
\omega _{0}^{2} &  c(\omega' ) \\
c(\omega ) & \omega ^{2}\,\delta (\,\omega - \omega'\,)
\end{array}
\right) \,, 
\end{equation}
written in matrix notation. The matrix element labels
$\omega $ and $\omega '$ are to be integrated over here.
What we are doing here is not
a diagonalization of the operator itself, which will be dealt with later.

The potential matrix may be decomposed into the two parts as
\( \:
{\cal V} = {\cal V}_{0} + {\cal V}' \,, 
\: \)
where ${\cal V}_{0}$ consists of the environment part alone, 
\begin{eqnarray}
&& 
{\cal V}_{0} = 
\left( \begin{array}{cc}
0 & 0 \\
0 & \omega ^{2}\,\delta (\omega - \omega')
\end{array}
\right) \,, 
\\ &&
{\cal V}' = 
|{\cal S}_{1}\rangle \langle {\cal S}_{1}| - 
|{\cal S}_{2}\rangle \langle {\cal S}_{2}|
\,, 
\\ &&
|{\cal S}_{1}\rangle = \left( \begin{array}{c}
\omega _{0}  \\
\frac{c(\omega )}{\omega _{0}}
\end{array}
\right) \,, \hspace{0.5cm} 
|{\cal S}_{2}\rangle = \left( \begin{array}{c}
0  \\
\frac{c(\omega )}{\omega _{0}}
\end{array}
\right) \,, 
\\ && 
\langle {\cal S}_{i}|{\cal S}_{i}\rangle  = 
\omega _{0}^{2}\,\delta _{i1} + \frac{1}{\omega _{0}^{2}}\,
\int_{\omega _{c}}^{\infty }\,d\omega \,2\omega \,r(\omega ) \,.
\end{eqnarray}
We may note a trivial relation:
\begin{equation}
|{\cal S}_{i}\rangle \langle {\cal S}_{i}| = 
\langle {\cal S}_{i}|{\cal S}_{i} \rangle \,{\cal P}_{i} 
\,, \hspace{0.5cm} {\cal P}_{i}^{2} = {\cal P}_{i} \,,
\end{equation}
where ${\cal P}_{i}$ is a projection operator onto a one-dimensional subspace.
The fact that the nontrivial part ${\cal V}'$ is finite dimensional
is the reason this system is solvable.

The eigenvector of diagonalized ${\cal V}$
is given with the aid of some analytic function.
First, one defines the proper self-energy $\overline{G}(z)$ and
the full propagator $F(z)$ by
\begin{eqnarray}
\overline{G}(z) &=& 
\int_{-\infty }^{\infty }\,\frac{d\omega }{2\pi }\,
\frac{r(\omega )}{z - \omega }  \,,
\\
F(z) &=& \frac{1}{-\,z^{2} + \omega _{0}^{2} + 2\pi 
\overline{G}(z)}
\,, 
\end{eqnarray}
using an extended response weight,
\( \:
r(- \omega ) = -\,r(\omega )
\: \)
for $\omega < -\,\omega _{c}$.
The function $F(z)$ has cuts along the real axis, 
\( \:
\omega > \omega _{c} 
\: \)
and 
\( \:
\omega < -\,\omega _{c} \,. 
\: \)
With the assumption of $\omega _{c} > 0$ there is a gap between the
two cuts.
The following discontinuity relation holds;
\begin{eqnarray}
F(\omega + i 0^{+}) - F(\omega - i 0^{+}) = i
2\pi \,r(\omega )F(\omega + i 0^{+})F(\omega - i 0^{+})
\equiv i 2\pi H(\omega ) \,, 
\end{eqnarray}
along this cut.
The frequency renormalization, corresponding to the mass renormalization
in field theory, is necessary when the high frequency behavior of the
response weight is
\( \:
r(\omega ) \rightarrow 
\: \)
constant.
The renormalized frequency is
\begin{equation}
\omega _{R}^{2} = \omega _{0}^{2} + \delta \omega ^{2} =
\omega _{0}^{2} - 2\,\int_{\omega _{c}}^{\infty }\,d\omega \frac{r(\omega )}
{\omega } \,. \label{renormalized frequency} 
\end{equation}
When this happens, one has to subtract a term in the continuous $\omega $
integral involving $r(\omega )$ and replace the bare $\omega _{0}^{2}$
by $\omega _{R}^{2}$. We shall not indicate this procedure
in further presentation, because it is fairly straightforward.

The eigenvector of ${\cal V}$ is then given by \cite{jmy-96-2}
\begin{eqnarray}
|\Psi (\omega )\rangle &=& 
-\, c(\omega )F(\omega  - i 0^{+})\,|0\rangle 
\nonumber 
\\ 
&+& |\omega \rangle 
+ c(\omega )F(\omega - i 0^{+})
\,\int_{\omega _{c}}^{\infty } d\omega' \frac{c(\omega' )}
{\omega'\,^{2} - \omega ^{2} + i 0^{+}}\,|\omega' \rangle 
\,. \label{out-state eq} 
\end{eqnarray}
The notation here is such that $|\omega \rangle$ and $|0\rangle $ 
are eigenvectors of ${\cal V}_{0}$:
\begin{eqnarray}
\left( \,{\cal V}_{0} - \omega ^{2}\,\right)\,|\omega \rangle = 0 \,, 
\hspace{0.5cm} 
{\cal V}_{0}\,|0\rangle = 0 \,.
\end{eqnarray}
This diagonalization involves a complex phase such as
\begin{eqnarray*}
e^{i\varphi (\omega )} = F(\omega - i 0^{+})/|F(\omega - i0^{+})| \,.
\end{eqnarray*}
Acutally one can show that these are overall phases of vectors and
by removing them one achieves diagonalization of the real symmetric matrix
${\cal V}$ by a real orthogonal transformation.

The matrix diagonalization precisely parallels the operator diagonalization.
The canonical transformation from the original to the Hamiltonian 
eigen operator is given by
\begin{eqnarray}
&& \hspace*{-1cm}
\tilde{Q}(\omega ) = Q(\omega ) - c(\omega )F(\omega - i 0^{+})\,
\left( \,q - 
\int_{\omega _{c}}^{\infty }\,d\omega '\,\frac{c(\omega ')}
{\omega '\,^{2} - \omega ^{2} + i 0^{+}}\,Q(\omega ')
\,\right) \,, 
\\ && \hspace*{-1cm}
\tilde{P}(\omega ) = P(\omega ) - c(\omega )F(\omega - i 0^{+})\,
\left( \,p - 
\int_{\omega _{c}}^{\infty }\,d\omega '\,\frac{c(\omega ')}
{\omega '\,^{2} - \omega ^{2} + i 0^{+}}\,P(\omega ')
\,\right) \,, 
\\ &&
q = -\,\int_{\omega _{c}}^{\infty }\,d\omega \,c(\omega )\,
F^{*}(\omega - i 0^{+})\,\tilde{Q}(\omega )  \,, 
\\ &&
p = -\,\int_{\omega _{c}}^{\infty }\,d\omega \,c(\omega )\,
F^{*}(\omega - i 0^{+})\,\tilde{P}(\omega ) \,, 
\\ &&
Q(\omega ) = \tilde{Q}(\omega ) + c(\omega )\,
\int_{\omega _{c}}^{\infty }\,d\omega' \,
\frac{c(\omega ')\,F^{*}(\omega' - i 0^{+})}
{\omega ^{2} - \omega '\,^{2} - i0^{+}}\,\tilde{Q}(\omega' ) \,, 
\\ &&
P(\omega ) = \tilde{P}(\omega ) + c(\omega )\,
\int_{\omega _{c}}^{\infty }\,d\omega' \,
\frac{c(\omega' )\,F^{*}(\omega' - i 0^{+})}
{\omega ^{2} - \omega '\,^{2} - i0^{+}}\,\tilde{P}(\omega' ) \,.
\end{eqnarray}
Here we find it more convenient to use 
\begin{eqnarray*}
\tilde{Q}(\omega ) \equiv e^{i\varphi (\omega )}\,\overline{Q}(\omega )
\end{eqnarray*}
instead of the true eigen operator $\overline{Q}(\omega )$,
by retaining those $\omega $ dependent phases.

Assuming the canonical commutation for the original variables,
one can verify that diagonal variables obey the correct form of
the commutation relation;
\begin{eqnarray*}
[ \,\overline{Q}(\omega ) \,, \overline{P}(\omega ')\,] =
i\,\delta (\,\omega - \omega '\,)
\end{eqnarray*}
etc.
It can be proved that with the specified phases of $\tilde{Q}(\omega ) \,, 
\tilde{P}(\omega )$ the original variables are all hermitian, as
required.

The overlap probability of the two vectors, or equivalently
the overlap between $q$ and $\overline{Q}(\omega )$, is 
\begin{eqnarray}
&&
|\,\langle 0|\Psi (\omega ) \rangle\,|^{2} =
c^{2}(\omega )\,|F(\omega - i0^{+})|^{2} = 2\omega \,H(\omega ) \,,
\\ &&
H(\omega ) =
\frac{r(\omega )}{(\,\omega ^{2} - \omega _{0}^{2}
- \Pi (\omega )\,)^{2} + (\pi r(\omega ))^{2}} \,, 
\end{eqnarray}
with 
\begin{equation}
\Pi (\omega ) = {\cal P}\,\int_{-\infty }^{\infty }\,\frac{d\omega '}{2\pi }\,
\frac{r(\omega ')}{\omega - \omega '} \,.
\end{equation}

The analytic function $F(z)$ contains the crucial information with regard
to the behavior of the small subsystem.
The system frequency $\omega _{0}$ originally given in the Lagrangian
is modified by the interaction with the environment.
The shifted frequency is determined by the singularity 
of the self-energy function $F(z)$.
With a condition,
\begin{eqnarray}
f(\omega _{c}^{2 -}) &>& \omega _{c}^{2} \,, 
\\
f(\lambda ) &\equiv&  \lambda + \Re F^{-1}(\sqrt{\lambda } + i0^{+}) \,,
\label{condition of no real pole} 
\end{eqnarray}
(in this equation $^{-}$ indicates the limit from below),
it can be shown that there is no singularity except the branch cut
starting from the threshold, 
\( \:
\omega > \omega _{c}
\: \)
and
\( \:
\omega < -\,\omega _{c} \,, 
\: \)
in the first Riemann sheet.
The original pole at $\omega = \omega _{0}$ thus moves into the second Riemann
sheet whose location is given by $z$ obeying
\begin{equation}
z ^{2} - \omega _{0}^{2} - 2\pi \,\overline{G}(z) 
+ 2\pi ir(z) = 0 \,. \label{spectrum zero} 
\end{equation}
The imaginary part of this location gives the decay rate of 
any initial configuration of the system, as will be made evident below.

\vspace{0.5cm} 
\begin{center}
\begin{large}
{\bf \lromn 4 Time evolution
}
\end{large}
\end{center}

With the diagonal variable derived,
it is easy to write the Heisenberg operator solution
at any time $t$ in terms of the initial operator values. For instance,
\begin{eqnarray}
&&
q(t) = -\,\int_{\omega _{c}}^{\infty }\,d\omega \,c(\omega )\,
F^{*}(\omega - i 0^{+})\,\tilde{Q}(\omega \,, t) \,, 
\\ &&
\tilde{Q}(\omega \,, t) = \cos (\omega t)\,\tilde{Q}_{i}(\omega ) +
\frac{\sin (\omega t)}{\omega }\,\tilde{P}_{i}(\omega ) \,.
\end{eqnarray}
The initial values $\tilde{Q}_{i} \,, \tilde{P}_{i}$ are then rewritten
in terms of the original variables.

After some straightforward calculation one finds \cite{jmy-96-2}  that
\begin{eqnarray}
q(t) &=& p_{i}\,g(t) + q_{i}\,\dot{g}(t) 
- \int_{\omega _{c}}^{\infty }\,
d\omega \,\sqrt{r(\omega )}\,\left( 
\,h^{*}(\omega \,, t)\,e^{-\,i\omega t}\,b_{i}(\omega ) + ({\rm h.c.})
\,\right) \,.
\label{q-correlator} 
\\ 
p(t) &=& 
p_{i}\,\dot{g}(t) + q_{i}\,\stackrel{..}{g}(t) 
- \int_{\omega _{c}}^{\infty }\,
d\omega \,\sqrt{r(\omega )}\,\left( 
\,k^{*}(\omega \,, t)\,e^{-\,i\omega t}\,b_{i}(\omega ) + ({\rm h.c.})
\,\right) \,. \nonumber \\ && \label{p-correlator} 
\end{eqnarray}
We have introduced
\begin{eqnarray}
g(t) &=& 2\,\int_{\omega _{c}}^{\infty }\,d\omega \,H(\omega )\,\sin (\omega t)
 \label{g-def} 
\\
&=&
2\,\int_{\omega _{c}}^{\infty }\,d\omega \,
\frac{r(\omega )\,\sin (\omega t)}{(\,\omega ^{2} - \omega _{0}^{2}
- \Pi (\omega )\,)^{2} + (\pi r(\omega ))^{2}} \,, 
\label{g-formula} \\
h(\omega \,, t) &=& \int_{0}^{t}\,d\tau \,g(\tau )\,e^{-\,i\omega \tau } \,, 
\\ 
k(\omega \,, t) &=& \int_{0}^{t}\,d\tau \,\dot{g}(\tau )e^{-\,i\omega 
\tau } = g(t)e^{- i\omega t} + i\omega h(\omega \,, t) \,.
\end{eqnarray}
Furthermore,
\( \:
b_{i}(\omega ) = (\,\sqrt{\omega }\,Q(\omega ) + i P(\omega )/
\sqrt{\omega }\,)/\sqrt{2}
\: \) is the annihilation operator for environment harmonic oscillators.

The important time dependence is governed by the function $g(t)$.
It is useful to clarify the physical significance of the function
$g(t)$ in detail.
For this purpose let us first derive a local form of the quantum Langevin
equaton.
Using the explicit form of solution, one gets after eliminating
initial $q_{i} \,, p_{i}$ dependence,
\begin{eqnarray}
&&
\frac{d^{2}q}{dt^{2}} + \Omega ^{2}(t)\,q + C(t)\,\frac{dq}{dt} =
-\,\Omega ^{2}(t)\,f_{q} - C(t)\,f_{p} - \dot{f}_{p} \,, 
\label{langevin local} 
\\ &&
\Omega ^{2}(t) = \frac{\dot{g}\stackrel{...}{g} - \stackrel{..}{g}^{2}}
{g\stackrel{..}{g} - \dot{g}^{2}} \,, \hspace{0.5cm} 
C(t) = \frac{\dot{g}\stackrel{..}{g} - g\stackrel{...}{g}}
{g\stackrel{..}{g} - \dot{g}^{2}} \,,
\\ &&
f_{q} =
\int_{\omega _{c}}^{\infty }\,
d\omega \,\sqrt{r(\omega )}\,\left( 
\,h^{*}(\omega \,, t)\,e^{-\,i\omega t}\,b_{i}(\omega ) + ({\rm h.c.})
\,\right) \,, 
\\ &&
f_{p} =
\int_{\omega _{c}}^{\infty }\,
d\omega \,\sqrt{r(\omega )}\,\left( 
\,k^{*}(\omega \,, t)\,e^{-\,i\omega t}\,b_{i}(\omega ) + ({\rm h.c.})
\,\right) \,.
\end{eqnarray}
The quantities that determine the subsystem behavior in this equation are 
the time dependent friction ($C(t)$) and
the time dependent frequency squared ($\Omega^{2}(t)$), which
incorporate environmental effects. 
Both of these are locally determined from $g(t)$.
Thus $g(t)$ describes an average behavior of the system variable
disregarding the random force from environment.
On the other hand, $f_{q}$ and $f_{p}$ give the random force from
the envrionment.

The system variable has been determined in terms of the initial
operator values of both the system and the environment variables.
Dependence on the system initial values $p_{i} \,, q_{i}$ are given by
the function $g(t)$. 
Both $g(t)$ and $\dot{g}(t)$ can be shown to obey
integro-differential equation of the following form ($y = g \;$ or
$\dot{g}$),
\begin{eqnarray}
&&
\frac{d^{2}y}{dt^{2}} + \omega _{0}^{2}\,y + 2\,\int_{0}^{t}\,d\tau \,
\alpha _{I}(t - \tau  )\,y(\tau ) = 0 \,, 
\label{g-equation} 
\\ &&
\alpha _{I}(\tau ) = -\,\frac{i}{2}\,\int_{-\infty }^{\infty }\,d\omega \,
r(\omega )\,e^{-\,i\omega \tau } =
-\,\int_{\omega _{c}}^{\infty }\,d\omega \,r(\omega )\sin (\omega \tau ) \,.
\label{noise kernel} 
\end{eqnarray}
The two functions, $g(t)$ and $\dot{g}(t)$, 
differ in their boundary conditions:
\( \:
g(0) = 0 \,, \; \dot{g}(0) = 1 \,.
\: \)

The characteristic behavior of the function $g(t)$ 
is that it decreases first
exponentially and then finally by an inverse power of time, as can be
seen in the following way.
Using the discontinuity formula, one may rewrite the $\omega $ integration
containing 
\begin{eqnarray}
H(\omega ) = \left( \,F(\omega + i 0^{+}) - F(\omega - i 0^{+})\,\right)
/ (2\pi i)
 \,, \nonumber 
\end{eqnarray}
along the real axis into the $F(z)$ integration, the
complex $z$ running both slightly above and below the cuts.
The factor $\sin (\omega t)$ is replaced by $\Im e^{-\,i\omega t}$
in this procedure.
A half of this complex contour can be deformed into the second sheet,
and one thereby encounters simple poles in the second sheet.
We assume for simplicity that there exists only a single pole 
in the nearby second sheet.
The intregral for $g(t)$ may then be expressed as 
the sum of the pole contribution (at $z = z_{0} $ with 
$ \Im z_{0} < 0$) in the second sheet and the contribution
parallel to the imaginary axis passing through $z = \omega _{c} $, both
in the first (\lromn 1) and in the second (\lromn 2)
sheet \cite{goldberger-watson}:
\begin{eqnarray}
&& 
g(\tau ) =
\Im \left( Ke^{-i\Re z_{0}\tau } \right)\,e^{\Im z_{0}\tau } 
\nonumber 
\\ && \hspace*{1cm} 
+ \,
\Im \left[ \frac{e^{i\omega _{c}\tau }}{\pi }\int_{0}^{\infty }dy\,
e^{- y\tau }\left( \,F_{{\rm \lromn 1}}(\omega _{c} + iy) - 
F_{{\rm \lromn 2}}(\omega _{c} + iy)\,\right) \right] \,, \label{g-integral} 
\end{eqnarray}
with
\( \:
K^{-1} = z_{0} - \pi \overline{G}'(z_{0} ) + i\pi r'(z_{0})\,.
\: \)

The pole contribution given by the first term describes
the exponential decay which usually lasts very long
during the most important phase of the decay period, while at very late times
the rest of contribution gives the power law decay.
In order to explain this late time behavior, let us take the response
weight,
\( \:
r(\omega ) = c\,(\omega - \omega _{c})^{\alpha } \,,
\: \)
for $\omega _{c} < \omega < \Omega $.
We assume the parameters in the range of 
$\Omega \gg \omega _{c}$. 
The late time behavior of $g(t)$ is derived from the continuous
part of $H(\omega )$ integration and is given by
\begin{eqnarray}
g(t) \approx \frac{2c}{\bar{\omega }^{4}}\,\Gamma (\alpha + 1)\,
\frac{\cos (\,\omega _{c}\,t + \frac{\pi }{2}\,\alpha \,)}
{t^{\alpha + 1}} \,,
\end{eqnarray}
where $\bar{\omega }$ is the pole mass assumed to obey
\( \:
\bar{\omega } \gg 
\: \)
the decay rate, the imaginary part of the pole location.
Thus, the power $-\,\alpha -1$ of 
\( \:
g(t) \propto t^{-\,\alpha - 1}
\: \)
is related to the threshold behavior of the response weight.

One can estimate the transient time $t_{*}$ from the exponential period to
the power period by equating the two formulas of $g(t)$ in their
respective ranges, to obtain
\begin{equation}
t_{*} \approx \frac{1}{\gamma }\,\ln \left( \frac{\bar{\omega }^{3}}
{2c\,\Gamma(\alpha + 1)\,\gamma ^{\alpha + 1}}\right)
 \,,
\end{equation}
with 
\( \:
\gamma = -\,\Im z_{0} \,
\: \)
the decay rate.
For a very small $c$ the factor inside the logarithm becomes large
($\propto c^{-\,2\alpha - 3}$), and by the time $t_{*}$ the initial
population has decreased like
\begin{equation}
e^{-\,2\gamma t_{*}} \:\propto  \: c^{4\alpha + 6} \,.
\end{equation}
It may thus be claimed that the power law behavior is difficult
to observe.
But we shall later show that this may not be so in cosmology.

We may call the approximation that neglects the non-pole contribution
for $g(t)$ as the resonance approximation since this approximation
is equivalent to taking a Breit-Wigner form for $H(\omega )$,
\begin{equation}
H(\omega ) \approx \frac{1}{\pi }\,\frac{\eta \omega }
{(\omega ^{2} - \omega _{R}^{2})^{2} + \eta ^{2}\omega ^{2}} \,, 
\end{equation}
and then integrating in Eq.(\ref{g-formula}) for $g(t)$ in the entire range
of $-\,\infty < \omega < \infty $ without considering the threshold effect.
It may also be called the local friction approximation 
since the equation (\ref{g-equation}) for $g(t)$ simply reduces to
\begin{equation}
\stackrel{..}{g} + \omega_{R}^{2}\,g + \eta \,\dot{g} = 0 
\end{equation}
in this case.
The solution of this equation is
\begin{eqnarray}
g(t) \approx
\frac{1}{\omega _{R}}\,\sin (\omega _{R}t)\,e^{-\,\frac{\eta }{2}\,t} \,,
\end{eqnarray}
for $\eta \ll \omega _{R}$.
This approximation is practically very useful in many cases, 
but there are physical effects such as the final
abundance of unstable particles in thermal medium that cannot be explained 
in this approximation.

A short-time behavior of quantum dissipation is neither well described
by the resonance approximation. We compare a precise estimate of
the time dependent friction $C(t)$ and the frequency $\Omega ^{2}(t)$
to the approximate one.
Assuming a smooth limit of the function $g(t)$, one has
\begin{eqnarray}
C(t) &\sim & -\frac{t^{3}}{3}\left( \,g^{(3)}(0)^{2} - g^{(5)}(0)\,\right)
\,, \\
\Omega ^{2}(t) &\sim & -\,g^{(3)}(0) \,, 
\end{eqnarray}
as $t \rightarrow 0^{+}$.
On the other hand, the pole approximation,
\( \:
g(t) \approx \frac{1}{\bar{\omega }}\,\sin (\bar{\omega }t)e^{-\,\gamma t}
\,, 
\: \)
gives 
\( \:
C(t) \approx 2\gamma \,, \hspace{0.3cm}
\Omega ^{2}(t) \approx \bar{\omega }^{2} + \gamma ^{2} \,.
\: \)

More seriously, the pole approximation violates the positivity of the
reduced density matrix:
\begin{equation}
\frac{d}{dt}\,{\rm tr}\;\rho ^{2} \approx  2\gamma \,{\rm tr}\;\rho ^{2}
\,.
\end{equation}
This relation implies that if the initial subsystem is in a pure quantum
state with
\( \:
{\rm tr}\;\rho ^{2} = 1\,, 
\: \)
it evolves into a state with
\( \:
{\rm tr}\;\rho ^{2} > 1 \,.
\: \)
Because 
\( \:
{\rm tr}\;\rho = 1 \,, 
\: \)
this cannot be satisfied unless some diagonal element of the density
matrix is negative, whose absolute value is larger than $1$.

A great advantage of the operator approach is that one may explicitly 
work out various correlators.
For instance,
\begin{eqnarray}
&&
\langle q(t_{1})q(t_{2}) \rangle = -\,\frac{i}{2}\,g(t_{1} - t_{2})
+ \int_{0}^{t_{1}}\,d\tau \int_{0}^{t_{2}}\,ds\,g(t_{1} - \tau )
\alpha _{R}(\tau - s)g(t_{2} - s) \nonumber 
\\ && \hspace*{-1cm}
+\, g(t_{1})g(t_{2})\,\langle p_{i}^{2} \rangle + \dot{g}(t_{1})
\dot{g}(t_{2})\,\langle q_{i}^{2} \rangle + (\,g(t_{1})\dot{g}(t_{2})
+ \dot{g}(t_{1})g(t_{2})\,)\,\frac{1}{2}\, \langle q_{i}p_{i} +
p_{i}q_{i}\rangle \,, 
\\ &&
\alpha _{R}(\tau ) = \int_{\omega _{c}}^{\infty }\,d\omega \,
\langle 2n_{i}(\omega ) + 1 \rangle\,\frac{\cos (\omega \tau )}{2\omega }\,.
\end{eqnarray}
Coincident time limits are evaluated from these, resulting in
\begin{eqnarray}
&& 
\langle q^{2}(t) \rangle = 
\int_{\omega _{c}}^{\infty }\,d\omega \,
\langle 2n_{i}(\omega ) + 1 \rangle\,r(\omega )\,|h(\omega \,, t)|^{2} 
\nonumber \\ &&
\hspace*{1.5cm} 
+ \,
g^{2}(t)\,\langle p_{i}^{2} \rangle + 
\dot{g}^{2}(t)\,\langle q_{i}^{2} \rangle + g(t)\dot{g}(t)\,
\langle p_{i}q_{i} + q_{i}p_{i} \rangle \,, 
\\ && 
\langle p^{2}(t) \rangle =
\int_{\omega _{c}}^{\infty }\,d\omega \,
\langle 2n_{i}(\omega ) + 1 \rangle\,r(\omega )\,|k(\omega \,, t)|^{2} 
\nonumber 
\\ &&
\hspace*{1.5cm} 
+\,
\dot{g}^{2}(t)\,\langle p_{i}^{2} \rangle 
+ \stackrel{..}{g}^{2}(t) \langle q_{i}^{2} \rangle + 
\dot{g}(t)\stackrel{..}{g}(t)
\langle p_{i}q_{i} + q_{i}p_{i} \rangle \,, 
\\ &&
\frac{1}{2}\, \langle q(t)p(t) + p(t)q(t) \rangle
= 
\int_{\omega _{c}}^{\infty }\,d\omega \,
\langle 2n_{i}(\omega ) + 1 \rangle\,r(\omega )\,h(\omega \,, t)
k^{*}(\omega \,, t) 
\nonumber \\ &&
+\,
\dot{g}(t)g(t)\,\langle p_{i}^{2} \rangle + \dot{g}(t)\stackrel{..}{g}(t)
\langle q_{i}^{2} \rangle + (\,\dot{g}^{2}(t) + g(t)\stackrel{..}{g}(t)\,)
\,\frac{1}{2}\, \langle p_{i}q_{i} + q_{i}p_{i} \rangle \,.
\end{eqnarray}
It is an important feature of these formulas that the initial state
dependence is clearly separated from the rest of physics as
\begin{eqnarray}
\langle\, 2n_{i}(\omega ) + 1 \,\rangle \,, \hspace{0.5cm} 
\langle q_{i}^{2} \rangle \,, \hspace{0.5cm} 
\langle p_{i}^{2} \rangle \,, \hspace{0.5cm} 
\langle p_{i}q_{i} + q_{i}p_{i} \rangle \,. \nonumber 
\end{eqnarray}

\vspace{0.5cm} 
\begin{center}
\begin{large}
{\bf \lromn 5 Relic abundance of unstable particles \\
in thermal medium
}
\end{large}
\end{center}

We shall apply the general result thus obtained to the unstable particle
decay. As already mentioned, in this context the system coordinate
$q_{\vec{k}}$ refers to the Fourier component of the field
operator of decaying particle $\varphi $.
In this section we discuss the number operator given by
\begin{equation}
n(t) \equiv \frac{1}{2}\, \langle \frac{p^{2}(t)}{\bar{\omega}} 
+ \bar{\omega}\,q^{2}(t) \rangle - \frac{1}{2} \,.
\end{equation}
Here the pole mass $\bar{\omega }$ is chosen as the reference frequency
to define the number operator from the Hamiltonian of the decaying field.

We first note that the asymptotic limit of the operator is given by
\begin{eqnarray}
&&
q(t) \:\rightarrow  \: -\,\int_{\omega _{c}}^{\infty }\,d\omega \,
\sqrt{r(\omega )}\,\left( \,F^{*}(\omega - i 0^{+})\,e^{-\,i\omega t}
\,b_{i}(\omega ) + ({\rm h.c.})\,\right) \,, 
\\ &&
p(t) \:\rightarrow  \: i\,\int_{\omega _{c}}^{\infty }\,d\omega \,\omega \,
\sqrt{r(\omega )}\,\left( \,F^{*}(\omega - i 0^{+})\,e^{-\,i\omega t}
\,b_{i}(\omega ) - ({\rm h.c.})\,\right) \,,
\end{eqnarray}
due to $h(\omega \,, \infty ) = F(\omega - i0^{+})$, which may readily
be proved.
Noting that
\begin{eqnarray}
&&
r(\omega )|h(\omega \,, \infty )|^{2} = H(\omega ) = 
\frac{|\,\langle 0|\Psi (\omega ) \rangle\,|^{2}}{2\omega } \,, 
\\ &&
r(\omega )|k(\omega \,, \infty )|^{2} = \omega^{2}H(\omega ) = 
\frac{\omega }{2}\, |\,\langle 0|\Psi (\omega ) \rangle\,|^{2} \,,
\end{eqnarray}
one has the asymptotic values,
\begin{eqnarray}
&&
\langle q^{2}(\infty ) \rangle = 
\int_{\omega _{c}}^{\infty }\,d\omega \,
\frac{|\,\langle 0|\Psi (\omega ) \rangle\,|^{2}}{2\omega }\,
\langle 2n_{i}(\omega ) + 1 \rangle \,, 
\\ &&
\langle p^{2}(\infty ) \rangle = 
\int_{\omega _{c}}^{\infty }\,d\omega \,\frac{\omega }{2}\,
|\,\langle 0|\Psi (\omega ) \rangle\,|^{2}\,
\langle 2n_{i}(\omega ) + 1 \rangle \,.
\end{eqnarray}
The overlap probability of the original subsystem variable with
the true eigen variable of the entire system
\( \:
|\,\langle 0 |\Psi (\omega ) \rangle\,|^{2}
\: \)
is thus fundamental to these and subsequent formulas.
The quantity
\( \:
\langle 2n_{i}(\omega ) + 1 \rangle
\: \)
refers to the number density of the environment bilinear fields.
We may take the value in thermal medium,
\begin{equation}
\langle 2n_{i}(\omega ) + 1 \rangle
= \coth (\frac{\beta \omega }{2}) \,.
\end{equation}

In the infinite time limit the occupation number is then
\begin{eqnarray}
n(\infty ) =
\frac{1}{2}\, \int_{\omega _{c}}^{\infty }\,d\omega \,\coth (\frac{\beta 
\omega }{2})\,(\bar{\omega} + \frac{\omega ^{2}}{\bar{\omega}})\,H(\omega )
- \frac{1}{2} \,.
\end{eqnarray}
When an precise form of the overlap
\( \:
H(\omega ) = \frac{|\,\langle 0|\Psi (\omega ) \rangle\,|^{2}}{2\omega }
\: \)
is used, this formula gives a reliable relic abundance of unstable
particles.

Let us first check that in some limit
this formula gives the familiar formula for the abundance.
When the pole term dominates, or equivalently one approximates $H(\omega )$
by the Breit-Wigner function,
then the temperature dependent part of the occupation number defined by
\begin{equation}
n^{\beta  } = n(\infty \,, \beta ) - n(\infty \,, 0) =
\int_{\omega _{c}}^{\infty }\,d\omega \,\frac{1}{e^{\beta \omega }
- 1}\,(\bar{\omega}  + \frac{\omega ^{2}}{\bar{\omega} })\,H(\omega ) \,, 
\end{equation}
has the factor 
\( \:
e^{-\,\bar{\omega }/T}
\: \)
at low temperatures.

In the high temperature regime the pole approximation is excellent.
But this approximation is not good at low temperatures.
Indeed, let us examine a typical example by taking again the form of
\( \:
r(\omega ) = c\,(\omega - \omega _{c})^{\alpha } \,, 
\: \)
with $0 < \alpha < 1$ in the range of $\omega _{c} < \omega < \Omega $
($\Omega \gg \omega _{c}$) and with
\( \:
\bar{\omega } \gg  {\rm Max}\;(\,\omega_{c} \,, T\,)
\: \).
The result is 
\begin{equation}
n^{\beta  } \approx \frac{c}{\bar{\omega }^{3}}\,\Gamma (\alpha + 1)\,
e^{-\,\beta \omega _{c}}\,T^{\alpha + 1} \,,
\end{equation}
where $\Gamma $ is the Euler's gamma function.
This shows that instead of the exponential suppression at low temperatures
what is left in medium after the decay has a power-law 
behavior of temperature dependence ($\propto T^{\alpha + 1}$).

An implication of this behavior to the unstable particle decay,
as will be made more explicit shortly,
is that the remnant fraction in thermal medium does not suffer from
the Boltzmann suppression factor at temperatures even much lower than
the mass of the unstable particle.
That this is possible is due to that the conventional approach using the
approximate Boltzmann-like equation is based 
on S-matrix elements computed on the mass shell, while the true
quantum mechanical equation may contain quantities off the mass shell.
As is well known, the Green's function, which is quantum mechanically
more fundamental than the S-matrix element,
does contain important contributions off the mass shell.
What is called virtual intermediate states in elementary quantum
mechanics gives rise to the off-shell contribution.
Exact treatment of the problem such as ours has indeed contributions off
the mass shell, and moreover the off-shell contribution makes up the
dominant part of behaviors at low temperatures.

Let us further apply these general 
considerations to the decay of unstable particle;
\( \:
\varphi \rightarrow \chi + \chi \,.
\: \)
Since we focus on the late time behavior,
the initial state dependence disappears: in particular, whether
the parent particle is or is not
in thermal equilibrium with the rest of medium is not important.
We shall limit our discussion here to the decay that occurs 
when the parent $\varphi $ becomes non-relativistic, 
\begin{equation}
\omega_{k} \sim  M + \frac{\vec{k}^{2}}{2M} \gg T \,,
\end{equation}
with $M$ the $\varphi $ mass.
This condition is relevant in interesting cosmological problems of the 
neutron decay at the time of nucleosynthesis and GUT $X$ boson decay at
baryogenesis.\cite{baryogenesis review96}

Since we already know the general expression for the relic abundance,
what remains to be done is the mode sum over the momentum $\vec{k}$ of the
unstable particle.
First, when the pole approximation is valid, the number density is
\begin{equation}
n \approx \int\,\frac{d^{3}k}{(2\pi )^{3}}\,e^{-\,\beta (\,M + k^{2}/2M\,)}
= (\frac{MT}{2\pi })^{3/2}\,e^{-M/T} \,.
\end{equation}
This is the familiar Boltzmann suppressed formula.

This is a bad approximation at low temperatures,
\( \:
T \ll M \,.
\: \)
To derive a precise formula, one has to integrate both over $\omega $
and $\vec{k}$, which is difficult to do analytically.
As an illustration, take a constant response weight
\( \:
r(\omega ) = r(\infty ) \,,
\: \)
hoping that the asymptotic region of $r(\omega )$ dominates.
Then,
\begin{eqnarray}
n &\approx& \frac{1}{2\pi ^{2}}\,\frac{r(\infty )}{M^{3}}\,
\int_{0}^{\infty }\,dk\,k^{2}\,
\int_{k}^{\infty }\,d\omega \,\frac{1}{e^{\omega /T} - 1}
\nonumber 
\\ 
&=& \frac{1}{90}\,\frac{r(\infty )\,T^{4}}{M^{3}} \,.
\end{eqnarray}
Although this is not a precise calculation, it nevertheless gives
a correct temperature dependence.

We numerically computed \cite{jmy-96-1} all terms including the logarithmic
factor in $r(\omega )$ along with $O[m^{2}]$ corrections.
It turns out that the total contribution is ten times larger 
than the analytic result above: in the $m\rightarrow 0$ limit,
\begin{equation}
n \approx  10^{-2}\,\frac{\mu ^{2}T^{4}}{M^{3}} \,.
\end{equation}
The main part of this large contribution comes from $|\omega | < k$.
With a dimensionless constant introduced by $\mu = gM$, 
this gives, relative to the photon number density
\( \:
( = \frac{2\zeta (3)}{\pi ^{2}}\,T^{3})
\: \),
\begin{equation}
\frac{n}{T^{3}} \approx  10^{-12}\,
(\,\frac{g}{G_{F}m_{N}^{2}}\,)^{2}\,\frac{T}{M} \,.
\end{equation}
We wrote the ratio here using the numerical value $G_{F}$,
the weak interaction constant of mass dimensions $-\,2$
(\( \:
G_{F}m_{N}^{2} \approx 10^{-\,5} 
\: \)).

One may estimate the equal time temperature $T_{{\rm eq}}$ 
at which the power contribution becomes equal
to the Boltzmann suppressed number density, to give
\begin{equation}
\frac{T_{{\rm eq}}}{M} \approx \frac{1}{30} \,, \hspace{0.5cm} 
\frac{n}{T^{3}_{{\rm eq}}} \approx 10^{-13} \,,
\end{equation}
taking as an example 
\( \:
\mu = 10^{-5}\,M \,,
\: \)
the weak interaction strength.
This number is in an interesting range that may affect nucleosynthesis, but
we should keep in mind that we did not work out the relevant three
body decay, 
\( \:
n \rightarrow p + e + \bar{\nu }_{e} \,.
\: \)

We shall mention another application of immediate interest in cosmology;
the heavy $X$ boson decay for GUT baryogenesis.
It has been argued \cite{baryogenesis review96} that there exists
a severe mass bound of order,
\begin{equation}
m_{X} > O[\alpha _{X}\,m_{{\rm pl}}] \approx 10^{16}\,{\rm GeV} \,, 
\label{x-mass const.} 
\end{equation}
to block the inverse process of the $X$ boson decay so that generation
of the baryon asymmetry proceeds with sufficient abundance of parent
$X$ particles.
The usual estimate of the mass bound mentioned above is however based
on the on-shell Boltzmann equation.
More appropriate formula in this estimate is our relic number density,
\begin{equation}
n_{X} \approx O[10^{-2}]\,g_{X}^{2}\,\frac{T^{4}}{m_{X}} \,.
\end{equation}
(In a more realistic estimate one should consider the $X$ boson decay
into quarks and leptons. But for an order of magnitude estimate difference
in statistics is not important.)
With the GUT coupling of $g_{X}^{2}/4\pi = 1/40$, the equal temperature
is roughly
\begin{equation}
T_{{\rm eq}} \approx \frac{M}{2} \,.
\end{equation}
Thus, already at temperature of about half of the $X$ mass the Boltzmann
suppressed formula is replaced by the power formula.
The kinematical condition Eq.(\ref{x-mass const.})
for baryogenesis must be reconsidered in view of our off-shell formula.

\vspace{0.5cm} 
\begin{center}
\begin{large}
{\bf \lromn 6 Path integral method
}
\end{large}
\end{center}

The basic idea of the influence functional method
\cite{feynman-vernon} is that one is interested in the
behavior of the subsystem alone
and traces out the environment variable altogether in the path
integral formula.
Furthermore, one directly deals with the probability instead of the amplitude.
This way one can compute the reduced density matrix 
that describes a state of the small system incorporating effects
of interacting environment.
We discuss this method here, simply because this approach gives
the non-decay probability of unstable particle
which is difficult to deal with in the operator approach so far used.

We define the influence functional by convoluting with the initial
state of the environment. To do so we assume for technical reasons
that initially we may take
an environment state uncorrelated with the system.
The influence functional is thus obtained after integrating 
out the environment variables:
\begin{eqnarray}
&&
{\cal F}[\,q(\tau )\,, q'(\tau )\,] \equiv 
\int\,{\cal D}Q(\tau )\,\int\,{\cal D}Q'(\tau )\,\int\,dQ_{i}\,\int\,dQ'_{i}
\int\,dQ_{f}\,\int\,dQ'_{f}\,
\nonumber \\
&& \hspace*{1cm} 
\cdot \delta (Q_{f} - Q'_{f})\,K \left(\,q(\tau )\,,Q(\tau ) \,\right)\,
K^{*} \left( \,q'(\tau )\,,Q'(\tau ) \,\right)\,\rho_{i} (Q_{i}\,, Q'_{i}) \,,
\\
&&
K \left( \,q(\tau )\,,Q(\tau ) \,\right) =
\exp \left( \,iS_{0}[Q] + iS_{{\rm int}}[q \,, Q]\,\right) \,, \\
&&
S_{0}[Q] + S_{{\rm int}}[q \,, Q] = \int_{0}^{t}\,d\tau \,
\left( \,L_{Q}[Q] + L _{{\rm int}}[q\,, Q]\,\right) \,.
\end{eqnarray}
The influence functional is a functional of the entire path 
of the system $q(\tau )$ and its conjugate path $q'(\tau )$.
\begin{equation}
\rho_{i} (Q_{i}\,, Q'_{i}) = \sum_{n}\,w_{n}\,\psi _{n}^{*}(Q_{i}\,')
\psi _{n}(Q_{i}) \,,  \hspace{0.5cm} (0 \leq w_{n} \leq 1) \,, 
\end{equation}
is the initial density matrix of the environment, which can be any
mixture of pure quantum state $n$ with the probability $w_{n}$.
What deserves to be stressed is that one does not observe the final
state of the environment, hence integration with respect to the final
values of $Q_{f} = Q'_{f}$ is performed here.

Once the influence functional is known, one may compute the transition
probability and any physical quantities of the $q-$system
by convoluting dynamics of the system under study. 
For instance, the transition probability is given, with introduction
of the density matrix $\rho ^{(R)}$, by
\begin{eqnarray}
&& 
\int\,dq_{f}\,\int\,dq'_{f}\,\psi_{f} ^{*}(q_{f})
\rho ^{(R)}(q_{f} \,, q'_{f})\,\psi_{f} (q'_{f}) \,, 
\\ && 
\rho ^{(R)} = 
\int\,{\cal D}q(\tau )\,\int\,{\cal D}q'(\tau )\,
\int\,dq_{i}\,\int\,dq'_{i}\,
\nonumber \\ && \hspace*{1cm} 
\cdot \psi_{i} ^{*}(q'_{i})\,\psi_{i} (q_{i})
\,{\cal F}[\,q(\tau )\,, q'(\tau )\,]\,
e^{iS_{q}[q] - iS_{q}[q']} \,, 
\end{eqnarray}
where $\psi_{i\,, f} $'s are wave functions of the initial and
the final $q-$states, and $S_{q}[q]$ is the action of the $q-$system.

The form of the influence functional is dictated by general principles
such as probability conservation and causality. Feynman and Vernon 
found a closed quadratic form consistent with these,
\begin{eqnarray}
{\cal F}[\,q(\tau )\,, q'(\tau )\,] &=& \nonumber \\
&& \hspace*{-3cm}
\exp \left[\,-\,\int_{0}^{t }\,d\tau \,\int_{0}^{\tau }\,ds\,
\left( \,\xi (\tau )\alpha_{R}(\tau - s)\xi (s) + i\,\xi (\tau )
\alpha _{I}(\tau - s)X(s)\,\right)\,\right] \,, 
\label{influence-f def} \\
&& \hspace*{-2cm}
{\rm with} \;
\xi (\tau ) = q(\tau ) - q'(\tau ) \,, \hspace{0.5cm} 
X(\tau ) = q(\tau ) + q'(\tau ) \,.
\end{eqnarray}
Thus two real functions $\alpha _{i}(\tau )$ are all we need to
characterize the system-environment interaction.
These are defined here in the range of 
\( \:
\tau \geq 0 \,.
\: \)
The fact that $\alpha_{i} $ depends on the difference of time
variables, $\tau - s$, is due to the assumed stationarity of the environment.
The Feyman-Vernon formula is valid for general $L_{Q}[Q]$ and
$L_{q}[q]$, not limited to the harmonic oscillator model if the interaction
$L_{{\rm int}}[q \,, Q]$ is bilinear.

The correlation kernels appear in the influence functional
as a form of the nonlocal interaction and they
are the dissipation $\alpha _{I}$ and the noise $\alpha _{R}$. 
The dissipation kernel $\alpha _{I}$ thus computed
agrees with the one defined in Eq.(\ref{noise kernel}).\cite{hjmy 96}
Let us now specialize to the case of the oscillator bath of temperature
$T = 1/\beta $,
which is described for a single oscillator of frequency $\omega $ by
\begin{eqnarray}
\rho _{\beta }(Q \,, Q') &=& \left( \frac{\omega }{\pi \,\coth (\beta \omega 
/2)}\right)^{1/2}\,
\nonumber \\ &&
\cdot \exp \left[ \,-\,\frac{\omega }{2 \sinh (\beta \omega )}
\,\left( \,(Q^{2} + Q'\,^{2})\,\cosh (\beta \omega ) - 2 Q Q'\,\right)\,\right]
\,. \label{thermal density matrix} 
\end{eqnarray}
The dissipation kernel is then
\begin{eqnarray}
\alpha _{R}(\tau ) = \frac{1}{2}\, \int_{-\infty }^{\infty }\,d\omega \,
\coth (\frac{\beta \omega }{2})\,r(\omega )\,e^{-\,i\omega \tau } \,.
\end{eqnarray}
Combined together, it gives the real-time thermal Green's function:
\begin{eqnarray}
&& \hspace*{-0.5cm}
\alpha (\tau ) \equiv \alpha _{R}(\tau ) + i\alpha _{I}(\tau )
= 
\sum_{k}\,c_{k}^{2}\,{\rm tr}\; \left( \,\rho _{\beta }\,T\left[ \,
Q(\omega _{k} \,, \tau )\,Q(\omega _{k} \,, 0)\,\right]\,\right) \,, 
\\ && \hspace*{-0.5cm}
\alpha (\omega ) \equiv \int_{-\infty }^{\infty }\,d\tau \,\alpha (\tau )
\,e^{i\omega \tau } =
i\,\sum_{k}\,c_{k}^{2}\,\left( \frac{1}{\omega^{2} -
\omega_{k}^{2} + i\epsilon } - \frac{2\pi i}{e^{\beta \omega _{k}} - 1}
\,\delta (\omega^{2} - \omega _{k}^{2}) \right) \,.
\nonumber \\ &&
\end{eqnarray}
As noted already, these are given in terms of the response weight
$r(\omega )$, and are governed by the analytic function $\overline{G}(z)$.

For the system dynamics we further assume a single harmonic oscillator
of frequency $\omega _{0}$.
In the path integral approach integration over the sum variable
$X(\tau )$ is trivial in this case,
since both the local part and the nonlocal action above
are linear in this variable:
\begin{eqnarray}
&&
\frac{i}{2}\, \int_{0}^{t}\,\left( \,\dot{\xi }(\tau )\dot{X}(\tau )
- \omega _{0}^{2}\,\xi (\tau )X(\tau )\,\right)  \nonumber 
\\ && \hspace*{-1cm}
- \,\int_{0}^{t }\,d\tau \,\int_{0}^{\tau }\,ds\,
\left( \,\xi (\tau )\alpha_{R}(\tau - s)\xi (s) + i\,\xi (\tau )
\alpha _{I}(\tau - s)X(s) \,\right) \,. 
\end{eqnarray}
Thus result of the path integration
of the system variable $X(\tau )$ gives the classical 
integro-differential equation for $\xi (\tau )$:
\begin{equation}
\frac{d^{2}\xi }{d\tau ^{2}} + \omega^{2} _{0}\,\xi (\tau ) +
2\,\int_{\tau }^{t}\,ds \,\xi (s)\,\alpha _{I}(s - \tau) = 0 \,.
\label{xi integro-eq} 
\end{equation}
The end result of the $\xi $ path integral then 
contains an integral of the form,
\begin{eqnarray}
-\,
\int_{0}^{t}\,d\tau \,\int_{0}^{\tau }\,ds\,\xi (\tau )\alpha _{R}(\tau - s)
\xi (s) \,,
\end{eqnarray}
using the classical solution $\xi (\tau )$ with specified boundary
conditions,
\( \:
\xi (0) = \xi _{i} \,, \hspace{0.3cm} \xi (t) = \xi _{f}
\: \).

In the local approximation often used the dissipation kernel 
is taken to have the form of
\begin{eqnarray}
\alpha _{I}(\tau ) = \delta \omega ^{2}\,\delta (\tau ) 
+ \eta \,\delta '(\tau ) \,, 
\end{eqnarray}
with $\delta \omega ^{2}$ representing 
the frequency shift and the $\eta $ term the local friction.
This choice enables one to solve the $\xi $ equation (\ref{xi integro-eq}) 
by elementary means.
On the other hand, the noise kernel is usually given by the response weight of
the form,
\begin{equation}
r(\omega ) = \frac{\eta }{\pi }\,\omega f(\frac{\omega }{\Omega }) \,, 
\end{equation}
with $f(x)$ some cutoff function and $\Omega $ a high frequency cutoff.
The cutoff is needed to tame the high frequency integral of 
$\alpha _{R}(t)$.
The simplest cutoff function 
\( \:
f(x) = \theta (1 - x)
\: \)
gives an approximate form of $\alpha _{I}(t)$ with the friction $\eta $
and
\begin{equation}
\delta \omega ^{2} \approx   -\,\frac{2}{\pi }\,\eta\,\Omega \,.
\end{equation}

The rest of deduction uses the Laplace transform, and we shall be brief,
leaving technical details to our original paper.\cite{jmy-96-1}
Solution of the integro-differential equation (\ref{xi integro-eq}) 
is, using $g(\tau )$ defined by eq.(\ref{g-def}), given as
\begin{eqnarray}
\xi (\tau ) =
\xi _{i}\,\frac{g(t - \tau )}{g(t)} + \xi _{f}\,
\left( \,\dot{g}(t - \tau ) - \frac{g(t - \tau )\dot{g}(t)}{g(t)} \,\right)
\,, 
\end{eqnarray}
with the dot denoting derivative.

The reduced density matrix of the quantum system at any time is obtained from
the action written in terms of the boundary values,
\( \:
S_{cl}(\xi _{f} \,, X_{f} \, ; \, \xi _{i} \,, X_{i}) \,, 
\: \)
by convoluting with the initial density matrix of the thermal environment.
This action is computed as
\begin{eqnarray}
i\,S_{{\rm cl}} &=& -\,\frac{U}{2}\,\xi _{f}^{2} - \frac{V}{2}\,\xi _{i}^{2}
- W\,\xi _{i}\,\xi _{f}  + 
\frac{i}{2}\,X_{f}\,\dot{\xi }_{f} - 
\frac{i}{2}\,X_{i}\,\dot{\xi }_{i} \,, 
\\
U &=& 2\,\int_{0}^{t }\,d\tau \,\int_{0}^{\tau }\,ds\,
z(\tau )\,\alpha _{R}(\tau - s)\,z(s) \,, 
\\
V &=& 2\,\int_{0}^{t }\,d\tau \,\int_{0}^{\tau }\,ds\,
y(\tau )\,\alpha _{R}(\tau - s)\,y(s) \,, 
\\
W &=& \int_{0}^{t }\,d\tau \,\int_{0}^{\tau }\,ds\,
\left( \,y(\tau )z(s) + y(s)z(\tau )\,\right)\,\alpha _{R}(\tau - s) \,, 
\\
y(\tau ) &=&
\frac{g(t - \tau )}{g(t)} \,, 
\\
z(\tau ) &=& \dot{g}(t - \tau ) - g(t - \tau )\frac{\dot{g}(t)}{g(t)} \,, 
\\
\dot{\xi }(\tau ) &=&
-\,\xi _{i}\frac{\dot{g}(t - \tau )}{g(t)} - \xi _{f}\,
\left( \,\stackrel{..}{g}(t - \tau ) - \frac{\dot{g}(t - \tau )\dot{g}(t)}
{g(t)}\,\right) \,.
\end{eqnarray}

For further discussion we take as the initial state 
a product of thermal states, a system of
temperature $T_{0} = 1/\beta _{0}$ 
and an environment of temperature $T = 1/\beta $.
We may take $T_{0} = T$ when we apply to the decay process of
excited level initially in thermal equilibrium.
On the other hand, 
in the limit of $T_{0} \rightarrow 0$ it describes the ground state
of the system harmonic oscillator.

After a series of straightforward
Gaussian integration we find the reduced density matrix as a function
of $X_{f}$ and $\xi_{f} $, of the form,
\begin{eqnarray}
&&
\rho ^{(R)}(X_{f}\,, \xi _{f}) = 2\sqrt{\frac{{\cal A}}{\pi }}
\exp [\,-{\cal A}X_{f}^{2} - {\cal B}\xi _{f}^{2} +
i{\cal C}\,X_{f}\xi _{f}\,] \,,
\label{density matrix} 
\\
&&
{\cal A} = \frac{1}{8 I_{1}}\,, \hspace{0.5cm} 
{\cal B} = \frac{1}{2}\, (\,I_{3} - \frac{I_{2}^{2}}{I_{1}}\,) \,, 
\hspace{0.5cm} 
{\cal C} = \frac{I_{2}}{2I_{1}} \,, 
\\
&&
I_{1} =
{\cal I}[\,|h(\omega \,, t)|^{2}\,] + 
\frac{1}{2\bar{\omega }}\coth (\frac{\beta _{0}\bar{\omega }}{2})\,
(\dot{g}^{2} + \bar{\omega }^{2}g^{2})
\,, 
\\
&&
I_{2} =
\Re {\cal I}[\,h(\omega \,, t)k^{*}(\omega \,, t)\,] +
\frac{1}{2\bar{\omega }}\coth (\frac{\beta _{0}\bar{\omega }}{2})\,
\dot{g}\,(\stackrel{..}{g} + \bar{\omega }^{2}g)
\,, 
\\
&&
I_{3} =
{\cal I}[\,|k(\omega \,, t)|^{2}\,]
+ \frac{1}{2\bar{\omega }}\coth (\frac{\beta _{0}\bar{\omega }}{2})\,
(\stackrel{..}{g}^{2} + \bar{\omega }^{2}\dot{g}^{2}) \,.
\label{i-def} 
\end{eqnarray}
Here $\bar{\omega } $ is a reference frequency taken 
as that of the initial system
state, and equated here to the frequency at the pole. 
If one so desires, either the renormalized 
$\omega _{R}$ or the bare $\omega _{0}$ may be taken as another
choice. If we imagine a situation in which the small system was added 
to a large environment at some time, 
its mutual interaction being absent prior to the initial time, then
it is appropriate to take $\omega _{0}$ as the reference frequency.
Since dependence on the initial state dies away quickly as time passes,
the choice of the initial reference is not crucial for determining
the behavior of states at late times.
Both of $h(\omega \,, t)$ and $k(\omega \,, t)$ are already
defined in the preceeding subsection.
The density matrix $\rho ^{(R)}$ from which any physical quantity at
time $t$
can be computed has explicitly been given by the discontinuity,
$H(\omega )$ or $r(\omega )$.

The basic quantities that appear in the reduced density matrix are
related to expectation values of the coordinate and the momentum
operators at the same moment by
\begin{eqnarray}
&&
\langle\, q^{2}\, \rangle = \frac{1}{8{\cal A}} = I_{1} \,, 
\\ &&
\langle\, p^{2}\, \rangle =
2{\cal B} + \frac{{\cal C}^{2}}{2{\cal A}} = I_{3} \,, 
\\ &&
\langle\, \frac{1}{2}\, (\,qp + pq \,)\, \rangle =
\frac{{\cal C}}{4{\cal A}} = I_{2} \,. 
\end{eqnarray}
Thus one may write the density matrix as
\begin{eqnarray}
&&
\rho ^{(R)}(X_{f}\,, \xi _{f}) =
\nonumber 
\\ && \hspace*{-1.5cm}
 \sqrt\frac{1}{2\pi\, \langle q^{2} \rangle}
\,\exp [\,- \,\frac{1}{8\langle q^{2} \rangle}\,X_{f}^{2} - 
\left( \,\frac{\langle p^{2} \rangle}{2} - 
\frac{\langle qp + pq \rangle^{2}}{8 \langle q^{2} \rangle}\,\right)
\,\xi _{f}^{2} + i\frac{\langle qp + pq \rangle}{4\langle q^{2} \rangle}
\,X_{f}\xi _{f}\,] \,.
\end{eqnarray}
The reduced density matrix is thus characterized by expectation
values of quadratic operators, just as in the case of pure Gaussian
system without the environmental effect.

It is sometimes useful to transform the density matrix in the
configuration space to the Wigner function $f_{W}(x\,, p)$,
\begin{eqnarray}
f_{W}(x \,, p) & \equiv& \int_{-\infty }^{\infty }\,d\xi \,
\rho^{(R)}(2x \,, \xi )\,e^{-\,i\,p \xi }
\\ &=&
\sqrt{\frac{4{\cal A}}{{\cal B}}}\,\exp [\,-\,4{\cal A}\,x^{2}
- \frac{(p - 2{\cal C}\,x)^{2}}{4{\cal B}}\,] \,.
\end{eqnarray}
The Wigner function is expected to give the probability distribution
in the phase space $(x\,, p)$ when the semi-classical picture is valid.
Expectation value of the number operator, namely the occupation number, 
in terms of the reference frequency,
equated to the pole location $\bar{\omega} $ here,
is calculated most easily from the Wigner function:
\begin{eqnarray}
\langle n \rangle &\equiv&  
\langle \,-\,\frac{1}{2\bar{\omega}}\,\frac{d^{2}}{dq^{2}} +
\frac{\bar{\omega}}{2}\,q^{2} - \frac{1}{2}\, \rangle 
= \frac{{\cal B}}{\bar{\omega}}
 + \frac{{\cal C}^{2}}{4\bar{\omega}\,{\cal A}}
 + \frac{\bar{\omega}}{16{\cal A}} - \frac{1}{2} 
\nonumber \\ 
&=&
\frac{1}{2\bar{\omega }}\,(\,I_{3} + \bar{\omega }^{2}
I_{1}\,) - \frac{1}{2}
\,. \label{number operator} 
\end{eqnarray}
It consists of two terms except the trivial $\frac{1}{2}$,  
the term $\bar{\omega}/(16\,{\cal A})$ from the Gaussian
width of the diagonal density matrix element and the rest from
the kinetic term $-\,\frac{d^{2}}{dq^{2}}$.
This formula of course agrees with that of the previous derivation in
the operator method.

\vspace{0.5cm} 
\begin{center}
\begin{large}
{\bf \lromn 7 Short-time behavior of non-decay probability
}
\end{large}
\end{center}

As an application of the influence functional method, I shall disucss
the short-time behavior of the decay probability.
This is an interesting problem from the point of the nucleon decay, as
mentioned in Introduction.
I shall describe some fundamental aspects of this problem, and leave
details to our forthcoming paper.\cite{jmy-97-1} 

We take as the initial state of the subsystem the first excited
state of harmonic oscillator that may be considered as a one-particle
state of unstable particle:
\begin{equation}
|i\rangle = a_{\bar{\omega }}^{\dag }|0\rangle \,, \hspace{0.5cm} 
\rho _{i} (q\,, q') = (\frac{\bar{\omega } }{\pi })^{1/2}\,2\bar{\omega }
\,qq'\,
e^{-\bar{\omega }(\,q^{2} + q'\,^{2}\,)/2} \,.
\end{equation}
After some straightforward computation one obtains the reduced density
matrix for this case:
\begin{eqnarray}
&& \hspace*{-1cm}
\rho ^{R} =
2\,\sqrt{\frac{{\cal A}}{\pi }}\,
(\,{\cal F}_{0} + {\cal F}_{1}X_{f}^{2} + {\cal F}_{2}\xi_{f} ^{2} + i
{\cal F}_{3}X_{f}\xi_{f} \,)\,
\exp [\,-\,{\cal A}\,X_{f}^{2} - {\cal B}\,\xi _{f}^{2} +
i\,{\cal C}\,X_{f}\,\xi _{f}\,] \,, \nonumber \\
&&
\\ &&
{\cal A} = \frac{1}{8 I_{1}}\,, \hspace{0.5cm} 
{\cal B} = \frac{1}{2}\, (\,I_{3} - \frac{I_{2}^{2}}{I_{1}}\,) \,, 
\hspace{0.5cm} 
{\cal C} = \frac{I_{2}}{2I_{1}} \,, 
\\ &&
{\cal F}_{0} =
\frac{1}{I_{1}}\,{\cal I}[\,|h(\omega \,, t)|^{2}\,] \,, \hspace{0.5cm} 
{\cal F}_{1} = \frac{1}{8\bar{\omega }I_{1}^{2}}\,
(\,\dot{g}^{2} + \bar{\omega }^{2}g^{2}\,) \,, 
\\ &&
\hspace*{-1cm}
{\cal F}_{2} =
-\,\frac{1}{2\bar{\omega }I_{1}^{2}}\,
\left( \,I_{1}^{2}\,(\,\stackrel{..}{g}^{2} + \bar{\omega }^{2}\dot{g}^{2}\,)
+ I_{2}^{2}\,(\,\dot{g}^{2} + \bar{\omega }^{2}g^{2}\,) -
2I_{1}I_{2}\,\dot{g}(\,\stackrel{..}{g} + \bar{\omega }^{2}g\,)\,\right)
\,,
\\ && \hspace*{-1cm}
{\cal F}_{3} =
\frac{1}{2\bar{\omega }I_{1}^{2}}\,
\left( \,{\cal I}[\,|h(\omega \,, t)|^{2}\,]\,\dot{g}(\,\stackrel{..}{g}
+ \bar{\omega }^{2}g\,) -
\Re {\cal I}[\,h(\omega \,, t)k(\omega \,, t)^{*}\,]\,
(\,\dot{g}^{2} + \bar{\omega }^{2}g^{2}\,)
\,\right) \,.
\nonumber \\ &&
\end{eqnarray}
The quantities $I_{i}$ are defined in Eq.(\ref{i-def}).

The non-decay probability, or sometime called the survival probability,
is defined as the overlap between this density matrix and that of the 
first excited state written by the final variables, $q_{f} \,, q_{f}'$.
This leads to
\begin{eqnarray}
&& 
\frac{\bar{\omega }}{4}\,\sqrt{{\cal A}\bar{\omega }}\,
\left( \,({\cal A} + \frac{\bar{\omega }}{4})({\cal B} 
+ \frac{\bar{\omega }}{4})
 + \frac{{\cal C}^{2}}{4}\,\right)^{-5/2}
\\
&& \hspace*{0.5cm} 
\cdot \left[ \,\left(\,({\cal A} + 
\frac{\bar{\omega }}{4})({\cal B} + \frac{\bar{\omega }}{4})
 + \frac{{\cal C}^{2}}{4}\, \right)\,
 \left( {\cal F}_{0}({\cal B} - {\cal A}) + {\cal F}_{1} - {\cal F}_{2}
 \right)  \right. \nonumber \\
&&
\hspace*{1cm} +
\left. \frac{3}{2}({\cal B} - {\cal A})\left( \,
{\cal F}_{1}({\cal B} + \frac{\bar{\omega }}{4}) + {\cal F}_{2}
({\cal A} + \frac{\bar{\omega }}{4}) - {\cal F}_{3}\frac{{\cal C}}{2}
\,\right)\,\right] \,.
\end{eqnarray}
The survival probability must be derived after deviding the disconnected
contribution that corresponds to the ground to the ground transition,
\begin{equation}
\left[ \,I_{1}I_{3} - I_{2}^{2} + \frac{1}{4}
+ \frac{\bar{\omega }}{2}\,I_{1} + \frac{I_{3}}{2\bar{\omega }}\,\right]
^{-1/2} \,.
\end{equation}
For the short-time behavior this gives a leading term for the
non-decay probability of the form,
\begin{eqnarray}
P_{1\rightarrow 1} \approx 1 - \frac{1}{\bar{\omega }}\,
\int_{\omega _{c}}^{\infty }\,d\omega \,
r(\omega )\coth (\frac{\beta \omega }{2})\,|k(\omega \,, t)|^{2} \,.
\end{eqnarray}
One is very much interested in whether the second term is proportional
to time$^{2}$ or not.\cite{maiani et al 93} 
Naively,
\begin{equation}
|k(\omega \,, t)|^{2} \:\rightarrow  \: \dot{g}(0)^{2}\,t^{2} =
t^{2} \,, 
\end{equation}
as $t \rightarrow 0^{+}$, but one must be careful, because the limit
may not be exchangeable with the $\omega $ integration when
the integral is conditionally convergent.

Since 
\begin{eqnarray*}
\coth (\frac{\beta \omega }{2}) = 1 + \frac{2}{e^{\beta \omega } - 1}
\,,
\end{eqnarray*}
the temperature dependent term is given by an absolutely convergent
$\omega $ integral, for which one may exchange the
\( \:
t \rightarrow 0^{+}
\: \)
limit and the integral. Thus, effect dependent on the environment gives
a term of order $t^{2}$ at short times.

On the other hand, the environment independent term differs, depending
on whether
\begin{eqnarray*}
\int_{\omega _{c}}^{\infty }\,d\omega \,r(\omega )
\end{eqnarray*}
is finite or not. If it is finite as is the case
in many applications in condensed
matter physics, the survival probability contains terms of order $t^{2}$.
But if it approaches a constant,
\( \:
r(\omega ) \rightarrow r(\infty ) \,, 
\: \)
then a more careful computation gives
\begin{equation}
P_{1 \rightarrow 1} \approx 1 - \frac{\pi r(\infty )}{\bar{\omega }}\,t
\,.
\end{equation}

The result thus seems to suggest that the short time limit in our boson
decay model gives the short time limit of the exponential decay,
\begin{equation}
e^{-\,\Gamma t} \approx 1 - \Gamma t \,.
\end{equation}
But, there are many details worth of further investigation which will be
dealt with elsewhere.\cite{jmy-97-1}

\hspace*{0.5cm} 
\begin{center}
{\bf Acknowledgments}
\end{center}

This work has been supported in part by the Grand-in-Aid for Science
Research from the Ministry of Education, Science and Culture of Japan,
No. 08640341.

\end{document}